# A novel simple and accurate flatness measurement method

*Thang H. L.*

*Vietnam Metrology Institute*

E-mail: thanglh@vmi.gov.vn

**Abstract** :

Flatness measurement of a surface plate is an intensive and old research topic. However ISO definition related and other measurement methods seem uneasy in measuring and/ or complicated in data analysis. Especially in reality, the mentioned methods don't take a clear and straightforward care on the inclining angle which is always included in any given flatness measurement. In this report a novel simple and accurate flatness measurement method was introduced to overcome this prevailing feature in the available methods. The mathematical modeling for this method was also presented making the underlying nature of the method transparent. The applying examples show consistent results.

## 1. Introduction

Flatness of a marble/granite plate is defined in the ISO 1101 [1] as the distance between the two tolerance planes surrounding the physical surface in such a way the two planes make the so called minimum zone. There have been several methods to realize this definition [2,3,4]. Several other methods were not really clear linking to the ISO definition but they are still accepted in practice [5 – 12]. However those methods share a common feature that they still include the inclining effect which is unavoidable in a given measurement. In other words, the inclining effect inherently sticks to the final flatness and it seems usually not small enough to be skipped.

We attempted to explore the root of the inclining effect in the existed methods by consideration of the measurement process and modeling it. The modeling equation shows a simple relationship and naturally drives us to measuring patterns or a measuring method in which the inclining effect cancellation could be automatically implemented.

## 2. Measurement modeling

The flatness Δ either following the ISO definition or not share the same equation of the following type:

$$\Delta = h_{max} - h_{min} \sim \Theta_{max} - \Theta_{min} \qquad (1)$$

Where, the subscripts max and min denote the points staying in the upper and lower planes respectively. The $\Theta$ means the measured angle and the h means the equivalent measured height. It is easy to recognize the relation between the angle $\Theta$ and the height h in case of the measurement tool is a level or an autocollimator as:

$$h = H \times \Theta \qquad (2)$$

Where, H is the length of the foot of the sliding mirror or level. In other method which may use a laser interferometer or a CMM, the angle may not be necessary shown explicitly, because the height h is often calculated implicitly inside the measuring system and so the height h is the final outcome of the system only.

Nevertheless implicitly or explicitly, the angle will be always there and the analysis will be very clear with this measurable quantity so we will start our analysis now by writing again the equation (1) in the form:

$$\Delta_a = \Theta_{max} - \Theta_{min}, \; \Theta_{max} \geq \Theta_{min} \qquad (3)$$

The following formula will be true by combining the equation (3) and the equation (2):

$$\Delta = H \times \Delta_a \qquad (4)$$

As a matter of fact that any angle measurement for an arbitrary point on a surface will be including at least the angle $\Theta_f$ due to the inherent uneven of the surface which defines the under question flatness and the inclining angle $\Theta_i$ which is the inclination of the table surface to a reference plane. Apparently, the just mentioned reference plane most conveniently chosen is a plane where all the points staying on it are the points which should have the same gravitational vector (in short, this plane will be called the g plane). Of cause this plane in turn is only a pseudo plane because it should have a finite curving radius. However this imperfection will be taken into the measurement uncertainty latter, so from now on it will be reasonably considered this reference plane as a perfect one. In fact this kind of reference plane will be convenient in the measurement using tools such as level or autocollimator.

The equation (3) can be rewritten in the following form:

$$\Delta_a = (\Theta_f + \Theta_i)_{max} - (\Theta_f + \Theta_i)_{min}$$

$$= (\Theta_{f\,max} - \Theta_{f\,min}) + (\Theta_{i\,max} - \Theta_{i\,min}) \qquad (5)$$

Equation (3) implies the flatness is a non negative quantity, therefore:

$$(\Theta_{f\,max} - \Theta_{f\,min}) \geq -(\Theta_{i\,max} - \Theta_{i\,min}) \qquad (6)$$

Let define:

$$\Delta_o = \Theta_{f\,max} - \Theta_{f\,min} \qquad (7)$$

Then the equation (5) now will be:

$$\Delta_a = \Delta_o + (\Theta_{i\,max} - \Theta_{i\,min}) \qquad (8)$$

It is necessary to note here a fact that a given plate is often inclining adjusted by a bubble level. The resolution of this kind of level or also the inclination adjustment itself often brings the plate to an angle of few micrometers to few tens of micrometers in height equivalence. This height equivalence is adversely in almost same order of magnitude of the flatness of a plate. This note means that the term $(\Theta_{i\,max} - \Theta_{i\,min})$ could get the numerical values arbitrarily depending on the inclining adjustment for the plate before the measurement. So it can be stated now that any measurement that does not count this effect may admit possible an arbitrary error in it.

## 3. Novel measuring patterns proposal

The equation (8) is very crucial because it shows us right away that the flatness we measured in general will always contain the components concerned the surface inclination which is apparently unavoidable in practical measurement. This equation will also tell us right away that in the ideal case where there is no inclination at all then the measured flatness $\Delta_a$ will be exactly equal to the $\Delta_o$ which can be now interpreted as the true flatness. This case is also equivalent to the case where the inclining angles of the considered max and min points are the same.

In order to cancel out this effect, a suitable method is the one should be able to make the inclination effect zero so that a true flatness $\Delta_o$ will be then obtained. It is easily to imagine that if we limit the measuring locus on a straight line only then all the points on this line will share a common inclining angle. Take an assumption that all the points on the other lines in parallel to each other will have to share also the same inclining angle. The following measuring patterns will be proposed as the one matched to determine the $\Delta_o$ as defined in (8) (**Figure 1**).

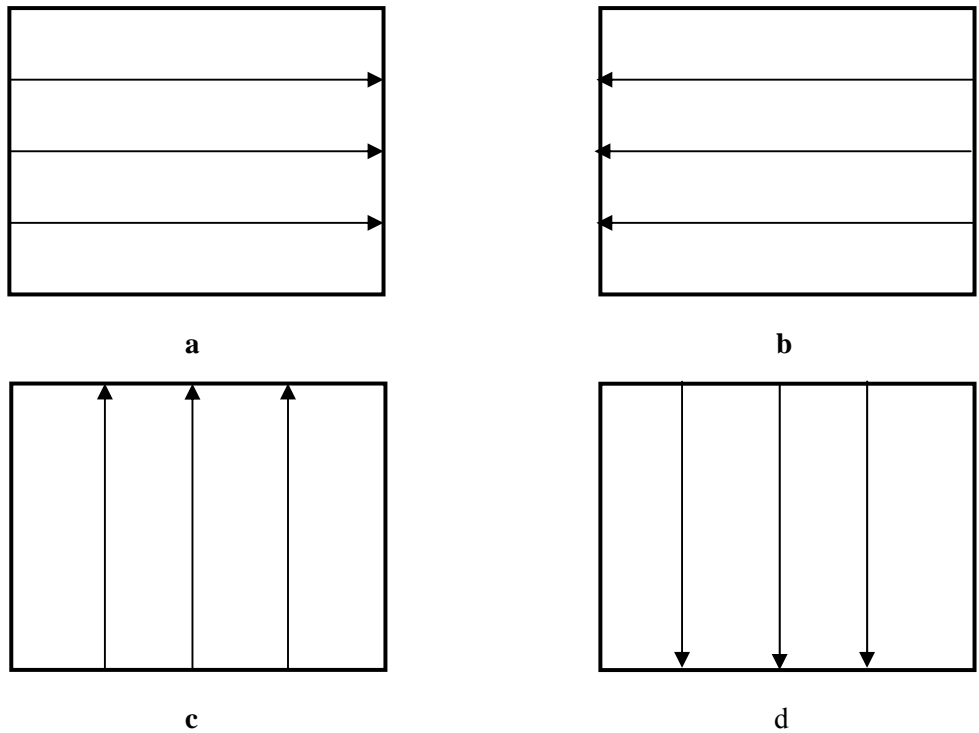

**Figure 1**. Proposed measuring patterns. The number of lines is case dependent.

The measurement following the proposed measuring patterns using a Wyler electronic level was carried out. This Wyler level was calibrated through the H.P. interferometer system which traced to the primary standard of length the $I_2$/He-Ne of Vietnam Metrology Institute (**Table 1**). The granite plate is made from the company Europe, Italia with dimension 1600×1000×180 mm, weight 900 Kg, accuracy 4.2 μm. The inclining angle of the plate was adjusted by a bubble level with the resolution 20 μm. As it can be seen latter the adjustment could bring the plate to an inclining angle of not larger than 18 μm in our experiment. A ruler was also used as a direction driving in the measurement. All of those tools were shown in **Figure 2**.

**Table 1**. Calibration results for the Wyler electronic level used as the measuring tool in this report. H = 150 mm. The calibration is traceable to S.I. meter.

| Level reading (μm) | Standard reading (μm) | Error = Level reading -Standard reading (μm) |
|---|---|---|
| -300 | -275.94 | -24.03 |
| -150 | -138.16 | -12.03 |
| -75 | -69.04 | -6.01 |
| -15 | -13.81 | -1.20 |
| -7.5 | -6.88 | -0.59 |
| -1.5 | -1.27 | -0.11 |
| -0.75 | -0.70 | -0.06 |
| 0 | 0 | 0 |
| 0.75 | 0.73 | 0.06 |
| 1.5 | 1.48 | 0.12 |
| 7.5 | 7.12 | 0.62 |
| 15 | 13.96 | 1.21 |
| 75 | 69.36 | 6.04 |
| 150 | 138.45 | 12.05 |
| 300 | 276.16 | 24.05 |

The calibration data of the electronic level will be used in the correction of the measurement. The outcomes from this experiment are summarized (**Table 2**). All the measurements in this report were implemented in the condition of temperature (20±1)°C, the relative humidity is (50±20)%.

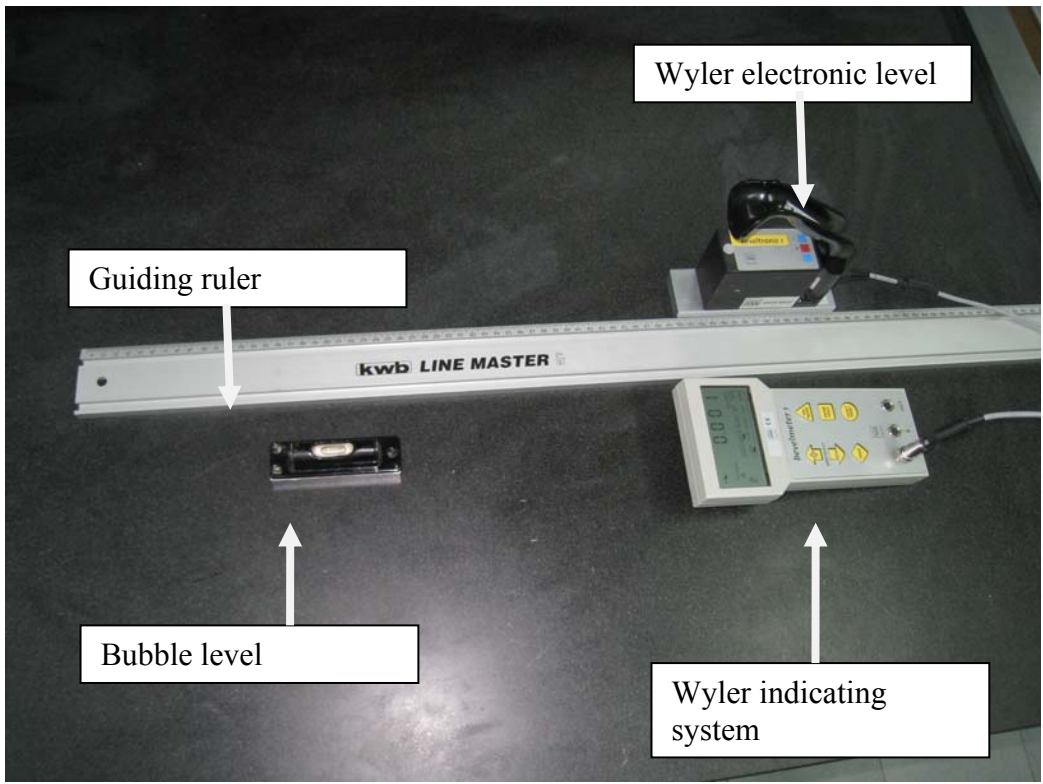

**Figure 2**. Photo of real measurement: the bubble level to adjust the plate inclination, the Wyler electronic level, the indicating system, the guiding ruler.

**Table 2**. Flatness following new proposed method

| Ord. | Method description | Measuring tool | Resultant flatness |
|---|---|---|---|
| 1 | Pattern Fig.1a, one way | Single Wyler electronic level | 1.4 μm |
| 2 | Pattern Fig.1b, one way | Single Wyler electronic level | 1.9 μm |
| 3 | Pattern Fig.1c, one way | Single Wyler electronic level | 1.0 μm |
| 4 | Pattern Fig.1d, one way | Single Wyler electronic level | 1.9 μm |

The uncertainty budget of the measurement was given in **Table 3**. All of those uncertainty components were followed the studies [13]. Although not all components contributed as dominantly as each other, they were still listed for the sake of completeness. The expanded measurement uncertainty estimated is 0.9 μm (k = 2, P = 95%). The measured flatness were corrected then plotted together with the estimated expanded uncertainties (**Figure 3**).

**Table 3**. Uncertainty Budget

| Influence Factor | Symbol | Distribution | Standard uncertainty mm | Contribution percentage % |
|---|---|---|---|---|
| 1. Measuring tool | $h_s$ | Gauss Type B | 1.00E-04 | 10.71 |
| 2. Earth curvature | $h_{td}$ | Chữ nhật Type B | 1.04E-06 | 0.00 |
| 3. Repeatability | $h_x$ | t (student) Type A | 0 | 0.00 |
| 4. Temperature gradient | $h_t$ | Chữ nhật Type B | 0 | 0.00 |
| 5. Supporting | $h_{cd}$ | Chữ nhật Type B | 2.33E-11 | 0.00 |
| 6. Closure error | $h_{ss}$ | Chữ nhật Type B | 0 | 0.00 |
| 7. Humidity | $h_{da}$ | Chữ nhật Type B | 0 | 0.00 |
| 8. Resolution | $h_r$ | Chữ nhật Type B | 2.89E-04 | 89.28 |

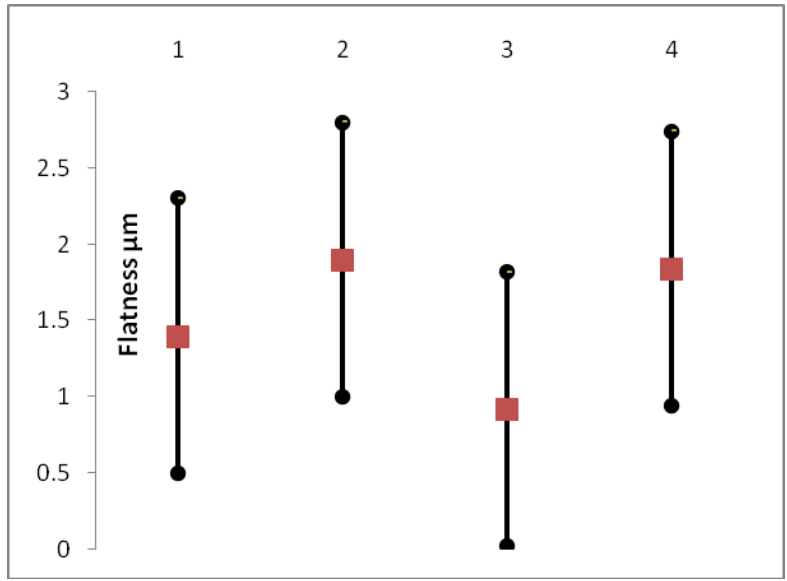

**Figure 3**. Flatness measurement results. The number 1, 2, 3 and 4 in the horizontal axis indicate the patterns as in **Table 1**.

### 4. Discussion

The inclining angle dependence of a given flatness measurement is apparent. This inclination dependence was able to be canceled in theory and they could be deleted by following our proposed measuring patterns.

The uncertainty estimated 0.9 μm is comparative to the other available ones [14]. From the uncertainty budget it is recognizable that the level resolution contribution was the largest one which occupies 89.28%. This fact will help to facilitate the uncertainty estimate in practice, since only this component consideration was enough in a routine flatness calibration work in our institute. However the full budget for sure was required in case we need something more accurately, but this will take a little bit more timing – cost.

The final measured results with the corresponding expanded uncertainties were plotted in the same graph show us that those measurements for the same plate with the same measuring instrument but with difference patterns providing consistent results. So the data process like the one in a key comparison will not be followed here. Last but not least, it is also understandable that the different patterns will bring to different unevenness picture of the same plate because each different pattern will give a differing view angle on the same object. This reflected the asymmetry of unevenness distribution of the plate. However the final flatness ones will stay consistent to each other as could be seen above.

Actually more different patterns could be possible if they were satisfactory to the condition that all the lines should share a common inclining angle. Therefore the patterns in our report here are not the only limited ones.

**5. Conclusion**

The inclination effect, a non ignorable effect in any given flatness measurement was addressed. The measurement modeling was done with a clear mathematical model. From this model, a measurement method was then introduced and measuring patterns were proposed.

Several experimental data following the proposed patterns were carried out and the corresponding uncertainties were estimated. Those results show different views of the same plate. The final results were consistent to each other. The proposed patterns were simple and able to make reaching the flatness true value.

**Acknowledgement**